\documentclass[12pt]{iopart}
\usepackage{times,bm,color}
\usepackage{amssymb,amsthm}
\usepackage{psfrag,epsfig}

\let\theta=\vartheta
\let\phi=\varphi
\let\rho=\varrho
\let\epsilon=\varepsilon
\let\setminus=\backslash
\def\d{{\rm d}}
\def\outer{{\rm box}}

\def\ie{i.e.~}

\def\n{{\rm n}}
\def\t{{\rm t}}
\def\g{{\rm g}}

\def\e{{\rm e}}
\def\L{{\rm L}}
\def\R{{\rm R}}
\def\IN{{\rm in}}
\def\OUT{{\rm out}}
\def\p{\;}
\def\real{{\mathbb R}}
\def\0{{}}
\def\eref#1{(\ref{#1})}

\def\Section{{section}}
\def\Sections{{sections}}
\def\Theorem{{theorem}}

\def\Proposition{{proposition}}

\def\Corollary{{corollary}}

\def\Lemma{{lemma}}

\def\Remark{{remark}}
\def\Remarks{{remarks}}
\def\Definition{{definition}}

\def\Fig{{figure}}
\def\Figs{{figures}}

\newtheorem{theorem}{Theorem}[section]
\newtheorem{proposition}[theorem]{Proposition}
\newtheorem{lemma}[theorem]{Lemma}

\newtheorem{corollary}[theorem]{Corollary}
\newtheorem{definition}[theorem]{Definition}
\newtheorem{remark}[theorem]{Remark}

\psfragscanon

\begin{document}

\title{Controllability for chains of dynamical scatterers}

\author{Jean-Pierre Eckmann$^{1,2}$ and Philippe Jacquet$^1$}

\address{$^1$ D\'{e}partement de Physique Th\'{e}orique, Universit\'{e} de
Gen\`{e}ve, CH-1211 Gen\`{e}ve 4, Switzerland}

\address{$^2$ Section de Math\'{e}matiques, Universit\'{e} de Gen\`{e}ve,
CH-1211 Gen\`{e}ve 4, Switzerland}

\ead{philippe.jacquet@physics.unige.ch}

\begin{abstract}

\noindent
In this paper, we consider a class of mechanical
models which consists of a linear chain of identical chaotic cells, each of which has
two small lateral holes and contains a rotating disk at its
center. Particles are injected at characteristic temperatures and rates from
stochastic heat baths located at both ends of the chain. Once in the system,
the
particles move freely within the cells and will experience elastic collisions
with the outer boundary of the
cells as well as with the disks. They do not interact with each other but
can transfer energy from one to another through collisions with
the disks. The state of
the system is defined by the positions and velocities of the particles and by the
angular positions and angular velocities of the disks. We show that each model
in this class is \emph{controllable} with respect to
the baths, \ie we prove that the action
of the baths can drive the system from any state to any other
state in a finite time. As a consequence, one obtains the existence of at most one
\emph{regular} invariant measure characterizing its states (out of equilibrium).
\end{abstract}

\vspace{-8mm}


\section{Introduction}

The study of heat conduction in (one-dimensional) solids remains a fascinating topic in theoretical physics. Various models have been developed to describe this phenomenon \cite{BLR,LLP}. In particular, the Lorentz gas model has been investigated and has been shown rigorously to satisfy Fourier's law \cite{LS}. However, since this model does not satisfy thermal local equilibrium (LTE) one cannot give a precise meaning to the temperature parameter involved in Fourier's law. To resolve this problem, a modified Lorentz
gas was proposed, where the scatterers (represented by disks) are still fixed in place but are now free to rotate \cite{LLM}. In this manner, the (non-interacting) particles can exchange energy from one to another through collisions with the scatterers. One clearly sees from numerical simulations that LTE is indeed satisfied and that heat conduction is accurately described by Fourier's law. To investigate such systems further, a class of models consisting of a chain of chaotic billiards, each containing a rotating scatterer, were introduced in \cite{EY2}. The authors developed a theory that
allows one, under physically reasonable assumptions (such as LTE), to derive rigorously
Fourier's law as well as
profiles for macroscopic quantities related to heat transport. They applied this theory to concrete examples that are either stochastic or deterministic. In
particular, they established a detailed analysis of a mechanical modified Lorentz
gas (MMLG) in which they assumed the existence and unicity of an
invariant measure describing its non-equilibrium steady state (\Section~4 in \cite{EY2}). To obtain a complete description of the MMLG model it thus remains to prove the existence and unicity of an invariant measure. While the question of existence is still a very challenging open problem, we shall show that there can be at most one \emph{regular} invariant measure.

In this
paper, we consider a class of mechanical models, extending the MMLG, and
show that every model in this class is \emph{controllable} with respect to
the baths, \ie  we prove that the action
of the baths can drive the system from any state to any other
state in a finite time. The result is formulated as \Theorem
\ref{main}, where we show that, starting from any initial state (comprising $n$ particles), the system can be emptied of any
particle, with all disks stopped at zero angular
position. The system being \emph{time-reversible}, this implies that one can fill it again with any number of particles and thus that one can drive the
system between any two states. (A set of states of zero Liouville measure has
to be excluded; this set consists of all states for which some particles stay forever in the system without hitting the disk or such that, in the course of time, will have simultaneous or tangential collisions with the disks or will realize corner collisions with the outer boundary of the cells.) As a consequence, one obtains for each model in the considered class, assuming the existence and enough regularity of an invariant measure characterizing its states (out of equilibrium), the \emph{uniqueness} of that invariant measure (see \Remark~\ref{unicity measure}). The
organization of this paper is as follows. In
\Sections~\ref{The Heat Baths} and \ref{Model} we present our assumptions on the baths and introduce the class of mechanical models considered.
Sections~\ref{One-cell analysis} and \ref{N-cell analysis} are
devoted to the controllability of the one-cell and $N$-cell
systems, respectively. In the conclusion we make some comments on possible generalizations.

\section{Heat baths}\label{The Heat Baths}

Although our discussion is mainly about the mechanical aspects of the
models, the notion of controllability is of course relative to
properties of the heat baths. Here, the exact details of the measure
describing the (stochastic) heat baths are not of importance. What counts
are only the sets of velocities and injection points into the
system. More precisely, we assume throughout the paper that, at any time, any open set of injection points
and velocities (including the direction) has \emph{positive measure}. In
particular, we shall exploit in a crucial way that any (open) set of realizations of the injection process with very high velocity indeed has positive measure. We shall use this positivity to inject ``driver'' particles to help emptying the system and thus obtain controllability as explained in the introduction.

\section{Mechanical models}\label{Model}

\begin{figure}[htbp]
\hspace{26mm}\psfig{file=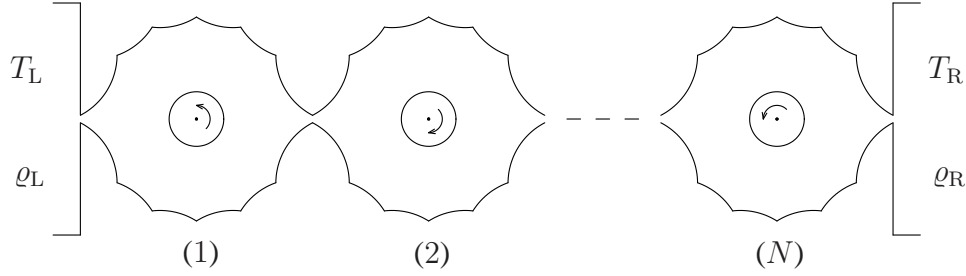,width=12cm,angle=180}
\vspace{3mm}\caption{The system composed of $N$ cells.}
\label{The System}
\end{figure}

The class of mechanical models considered in this paper consists of a linear
chain of
identical chaotic cells, each of which has two small lateral holes and contains a
rotating disk at its center (see \Fig~\ref{The System}). Particles are
injected at characteristic temperatures $T_{\L}$, $T_{\R}$ and rates
$\rho_{\L}$, $\rho_{\R}$ from
stochastic heat baths located at both ends of the chain (see \Section~\ref{The Heat Baths}). Once in the system,
the particles move freely within the cells and will experience elastic collisions
with the outer boundary of
the cells as well as with the disks. They do not interact with each
other but can exchange energy through collisions with
the disks. The state of
the system is defined by the positions and velocities of the particles and by the
angular positions and angular velocities of the disks. We will give a more
precise definition of phase space in \Section~\ref{dyn}.

We next specify the dynamics of the system (composed of $N$ cells) in more detail:
When there are $n$ particles in the system, we number them
as $i=1,\dots,n$ and denote by $q_1,\dots,q_n$ and $v_1,\dots,v_n$ their positions and velocities, respectively. Their trajectories are made of straight line segments joined at
the outer boundary of the cells or at the boundary of the disks. If a
particle reaches one of the two openings $\partial\Gamma^{(1)}_{\L}$ or
$\partial \Gamma^{(N)}_{\R}$, it leaves the system (and the remaining
particles are arbitrarily renumbered). Particles are injected into the system (from the baths) through
these boundary pieces as well. We write $\omega _1,\dots,\omega _N$ for the
angular velocities of the disks and $\varphi_j$ for the
angle a marked point on the rim of disk $j$ makes with the horizontal line passing through the center of disk $j$ ($j=1,\dots,N$). 

To describe the rules of the dynamics, let us focus on one of the $N$ cells, say the $j$th cell $\Gamma=\Gamma^{(j)}$, and assume that $q_i \in \partial \Gamma$ for some $1 \leq i \leq n$. We denote by $D$ the disk at the center of $\Gamma$, by $\partial \Gamma_\outer$ the outer boundary of $\Gamma$ and by $\partial
\Gamma_{\L}$ and $\partial \Gamma_{\R}$ its openings; they are either exits to the adjacent cells or to the heat baths. For a piecewise regular boundary $\partial
\Gamma = \partial \Gamma_\outer \cup \partial D$, there are unit vectors $\e_{\n}$ and $\e_{\t}$,
respectively normal outwards and tangent to $\partial \Gamma$ at $q_i$, and one
can write $v_i =
v_i^{\n} \e_\n + v_i^{\t} \e_\t$. We assume that the particles collide specularly
from the boundary
$\partial\Gamma_\outer  \setminus  (\partial \Gamma_{\L} \cup \partial \Gamma_{\R})$
and that the collisions between the particles
and the disk are elastic, so that for appropriate values
of the parameters (\ie the mass of the particles, the mass and the radius of
the disk), one obtains the following dynamical rules, where primes denote
the
values after the collision:
\begin{enumerate}
\item[1.] If $q_i \in \partial \Gamma_{\L} \cup
\partial \Gamma_{\R}$, then the $i$th particle keeps moving in a straight line to the adjacent cell or leaves the system.
\item[2.] If $q_i \in \partial \Gamma_\outer  \setminus  (\partial \Gamma_{\L} \cup
\partial \Gamma_{\R})$, then
\begin{equation}\label{Rule 1}
(v_i^{\n})' = -v_i^{\n}\p, \hspace{10mm} (v_i^{\t})' = v_i^{\t}\p.
\end{equation}
\item[3.] If $q_i \in \partial D$, then
\begin{equation}\label{Rule 2}
(v_i^{\n})' = -v_i^{\n}\p, \hspace{10mm} (v_i^{\t})' = \omega\p,  \hspace{10mm}
\omega' =
v_i^{\t}\p.
\end{equation}
\end{enumerate}
The position of the $i$th particle and the angular position of the disk after the collision are left unchanged.


\subsection{Geometry of the cell}\label{Geometry of a Cell}

\begin{figure}[htbp]
\hspace{27mm}\psfig{file=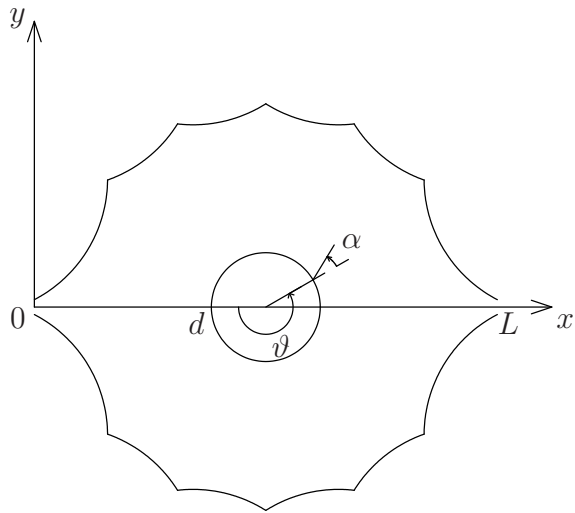,width=7cm}
\caption{A typical cell.}
\label{The Cell}
\end{figure}

In this subsection we describe the class of cells for which we can
prove controllability. Our definition is a compromise between
generality and tractability. In particular, this definition will allow
for a relatively simple controllability strategy. The reader who wants
to proceed to the controllability can just look at \Fig~\ref{The Cell} and use
that example as a typical cell.

Let $\Gamma_{\outer}$ be a bounded connected closed domain in $\real^2$ and let $L$ denote its width, that is
$(x,y)\in
\Gamma_{\outer} $ implies $x\in[0,L]$. We assume 

\begin{enumerate}
\item[1.] The boundary $\partial\Gamma_\outer$
of $\Gamma_{\outer}$ is made of two straight segments (the ``openings'') and a finite
number of arcs of circle, \ie
\begin{equation}
\partial \Gamma_\outer  = \partial \Gamma_\L \bigcup \partial \Gamma_\R \bigcup\left
(\bigcup_{k = 1}^{b} \partial \Gamma_{k}\right)\p,
\end{equation}
where  $\partial\Gamma_{\L} = \{(0,y) \ | \ y \in [-a,a]\}$,
$\partial\Gamma_{\R} = \{(L,y) \ | \ y \in [-a,a]\}$ ($2a$ corresponds to the
size of the openings) and each $\partial \Gamma_{k}$ is an arc of
circle. The arcs of circle are oriented so that $\partial\Gamma_{\outer}$ is everywhere dispersing (see \Fig~\ref{The Cell}).  

\item[2.] In
the interior of $\Gamma_{\outer}$ lies a disk $D$ of center $c=(L/2,0)$
and radius $r$. The disk does not intersect the boundary of $\Gamma_{\outer}$, \ie $\partial D \cap \partial \Gamma_\outer = \emptyset$.
\item[3.] Every ray from the center of the disk intersects the boundary
  $\partial\Gamma_\outer  $ only once: For every $z \in \partial \Gamma_\outer $ the segment $[c,z]$ intersects
$\partial\Gamma_\outer $ only at $z$, \ie $[c,z] \cap \partial\Gamma_\outer  =z$.
\end{enumerate}

\begin{definition}
  The closed domain $\Gamma = \Gamma_{\outer} \setminus D$ (with boundary $\partial
  \Gamma =
\partial \Gamma_{\outer} \cup
\partial D$) is called a \emph{cell}.
\end{definition}

\begin{figure}[htbp]
\hspace{25mm}\psfig{file=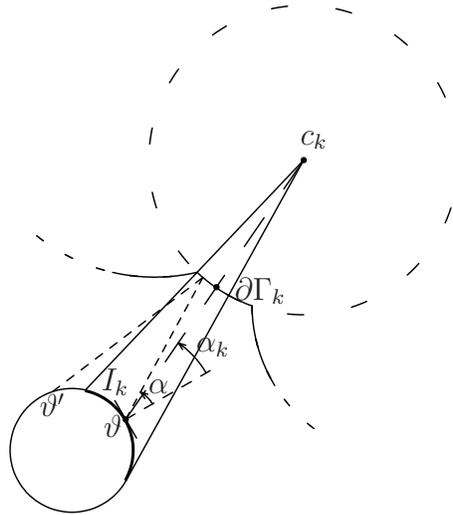,width=6cm}
\caption{The illumination construction: The illuminated segment $I_k$ is the part of
the disk (thick line) delimited by the two rays arising from $c_k$ and
going through the $k$th arc $\partial\Gamma_k$. The return path
corresponding to $R : (\theta,\alpha) \mapsto \theta'$ is shown as a dashed line.}
\label{The Shadows Construction}
\end{figure}

Our construction of $\partial\Gamma_\outer $ is motivated by the study of the
\emph{return map} $R$ from the disk to the disk under the dynamics of
the particle (see \Fig~\ref{The Shadows Construction}). We parameterize the
points on $\partial
D$ by the angle $\theta\in[0,2\pi)$ and denote by $\alpha
  \in [-\frac{\pi}{2},\frac{\pi}{2}]$ the angle a line makes with the outward
normal to the
  circle at $\theta$ (see \Fig~\ref{The Cell}). The return map $R$ is
  defined for $(\theta,\alpha)$ satisfying the following property: When a
  particle leaves the disk from $\theta$ in the direction 
$\alpha$, it returns
to the disk after {\em one} collision with the boundary
  $\partial \Gamma_\outer$ (and lands at $\theta'$). In that case, we define
$R(\theta,\alpha) = \theta'$. For other values of
$(\theta,\alpha)$, we say that $R$ is undefined. The domain of $R$ obviously depends
on the boundary $\partial \Gamma_\outer$. 

We next narrow the construction of acceptable domains by introducing the notion of 
{\em illumination}.
For each $k \in \{1,\dots,b\}$, we denote by $I_k$ the set of $\theta$
for which $R(\theta,\alpha_k(\theta))=\theta$ for some value $\alpha_k(\theta)$ of
$\alpha$ and so that
the reflection occurs on $\partial \Gamma_k$. Since the collisions with the corner points of $\partial \Gamma_k$ are undefined and the 
line connecting the boundary points of $I_k$ to the center $c_k$ (see \Fig~\ref{The Shadows Construction}) may be tangent to the disk, we actually neglect the boundary points of $I_k$, \ie we define $I_k$ as the largest \emph{open} (connected) set satisfying the above criteria.

\begin{figure}[htbp]
\centerline{\psfig{file=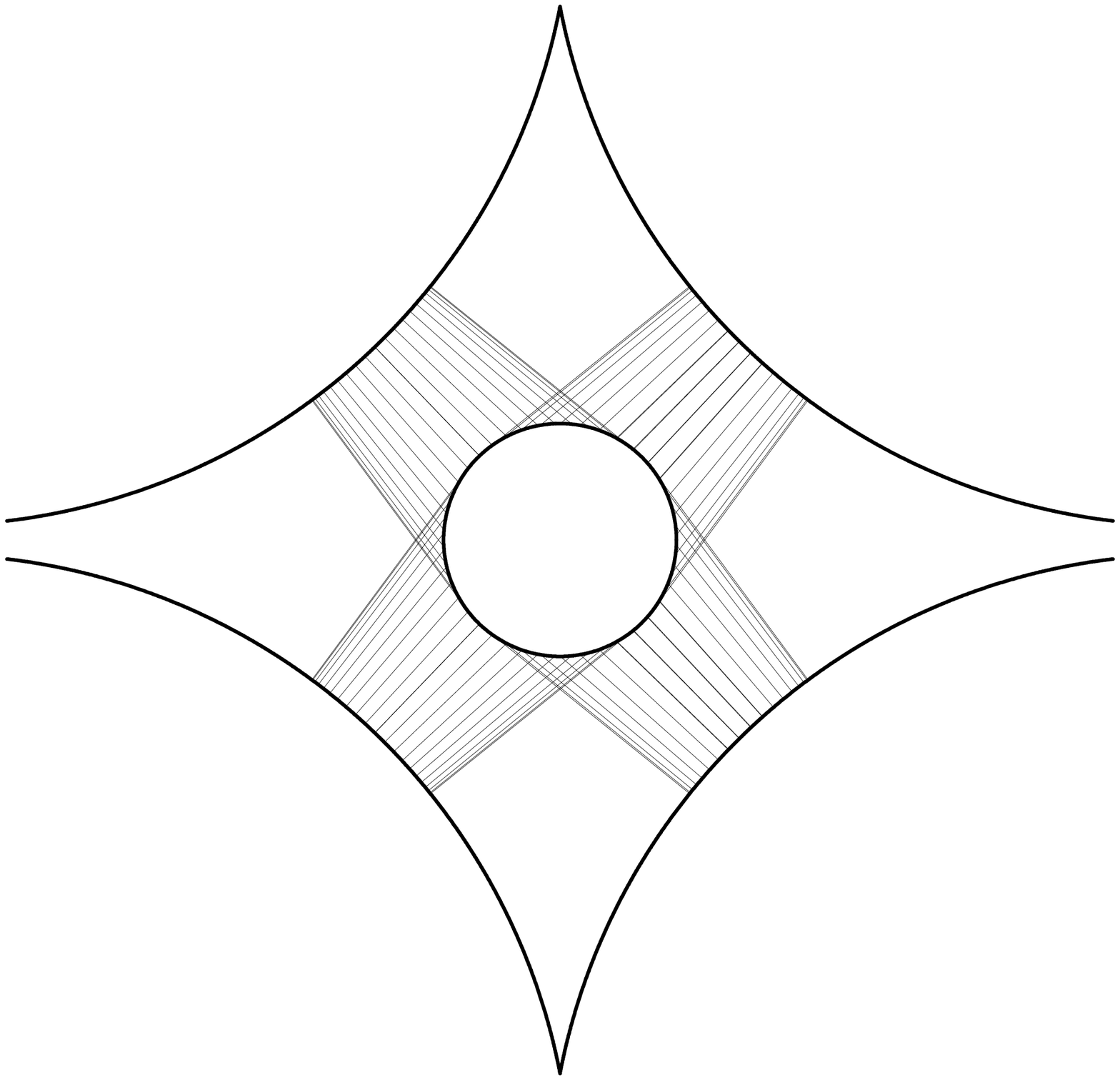,height=4.8cm} \hspace{1cm}
\psfig{file=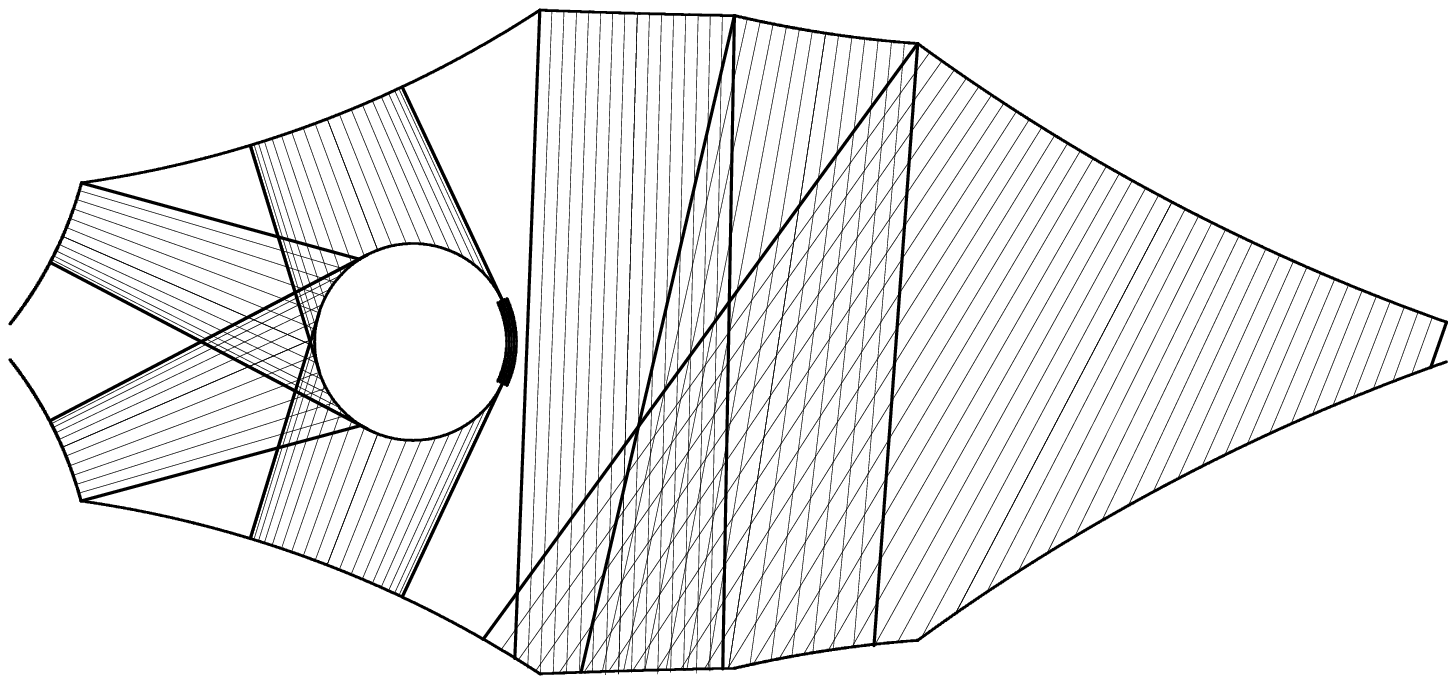,height=4.35cm}}
 \caption{The illuminated segments are the parts of the disk delimited
   by the outermost pairs of rays emanating perpendicularly
from the arcs. Left: A $1$-controllable cell. Right: This cell is not
$1$-controllable since the illuminations do not cover the part of the disk shown in thick
line.  The illuminations on the right are shown for the arcs on the
   top only. Basically, domains with long
``tails'' will not be $1$-controllable.}
\label{Not 1 Contrallable}
\end{figure}

This set can be more easily understood as follows:
Let  $C_k$ be the circle on which $\partial \Gamma_k$
lies and
let $c_k\in\real^2$ be its center.
If we ``shine'' light from that center to the disk, with only the $k$th arc $\partial \Gamma_k$
letting the light go through, then $I_k$ is in
fact that portion of the boundary of the disk on which light shines
from $c_k$ (and $\alpha_k (\theta)$ is the direction pointing
from $\theta$ to the center~$c_k$). Thus, $I_k$ is illuminated from
$c_k$. See \Figs~\ref{The Shadows Construction} and \ref{Not 1 Contrallable}.

\begin{remark}
 \textnormal{Notice that if a particle leaves the disk at
$\theta$ in the direction $\alpha$ and hits the $k$th arc $\partial
\Gamma_k$, then $R(\theta,\alpha) > \theta$  if
$\alpha > \alpha_k(\theta)$ and $R(\theta,\alpha) < \theta$ if $\alpha <
\alpha_k(\theta)$; see \Fig~\ref{The Shadows Construction}. In other
words, the return map $R$ maps {\em away} from the line pointing to the
center $c_k$.}
\end{remark}

\begin{remark}
 \textnormal{Notice that the illuminated segments $I_1,\dots,I_b$ will in general overlap.}
\end{remark}

\begin{definition}\label{Assumptions Cell}
A cell is called \emph{1-controllable} if the illuminated segments cover the entire boundary
of the disk, \ie
\begin{equation*}
\bigcup_{k=1}^b I_k =\partial D\p~.
\end{equation*}
\end{definition}

\begin{remark}
   \textnormal{We chose the term 1-controllable because our
  controllability proof will involve exactly one
  collision with $\partial \Gamma_\outer $ between any two consecutive collisions with the disk. One can imagine controllability
  proofs for domains with returns to the disk after several collisions
  with $\partial \Gamma_\outer $, and this would allow for more general
  domains. However, the gain of generality is perhaps not worth the effort.}
\end{remark}

\begin{remark}
 \textnormal{Note
that, since the illuminated regions $I_1,\dots,I_b$ are \emph{open} sets, one needs at least three generating circles
to make a 1-controllable cell. There are domains which are \emph{not} 1-controllable. See \Fig~\ref{Not 1 Contrallable}}.
\end{remark}

\subsection{Phase space}\label{dyn}

We next turn to the characterization of the phase space of the system
consisting of one cell and an arbitrary number of particles. We denote
by
\begin{equation}\label{State Space with n particles}
\Omega_{n} = (\Gamma^{n} \times [0,2\pi) \times
\real^{2n+1})/\sim
\end{equation}
the state space with $n$ particles, where $\mathbf{q} = (q_{1},\dots,q_n) \in
\Gamma^n$ denotes the positions of the
$n$ particles, $\phi \in [0,2\pi)$ denotes the angular
position of a (marked) point on the boundary of the turning disk, $\mathbf{v} =
(v_{1},\dots,v_n) \in
\real^{2n}$ denotes the velocities of the $n$ particles, $\omega \in
\real$  denotes the angular velocity of the turning disk (measured in the clockwise direction), and
$\sim$ is the
relation that identifies pairs of points in the collision manifold
$M_{n}=\{(\mathbf{q},\phi,\mathbf{v},\omega) \ | \ q_{i} \in \partial \Gamma
\mbox{ for some }
i\}$.

The phase space of the system (for one cell) is
\begin{equation*}
\Omega = \bigcup_{n=0}^\infty  \Omega_n \hspace{1cm} \mbox{(disjoint union)}~,
\end{equation*}
where now $n$ is the current number of particles in the cell. When a particle
is injected into the cell, the state of the system changes from $\xi\in
\Omega_n$ to a state in $\Omega_{n+1}$ obtained by adding to  $\xi$
a particle with position $q_{n+1}\in\partial\Gamma_L \cup
\partial\Gamma_R$ and velocity $v_{n+1}\in\real^2$ pointing
into the cell. Similarly, when a particle
leaves the cell, the corresponding two coordinates $q_i$ and $v_i$ are dropped.
We refer to \cite{EY2} for a detailed discussion of the numbering of the particles.

We denote by $\Phi_n^{t}$ the flow on $\Omega_n$. As long
as no collisions are involved, we have
\begin{equation}\label{Flow n}
\Phi_n^{t}(\mathbf{q}, \phi, \mathbf{v}, \omega) = (\mathbf{q}+\mathbf{v}t,
\phi +  \omega t
\kern -0.6em\pmod{2\pi}, \mathbf{v},
\omega)\p~.
\end{equation}
Clearly, if one specifies a realization $\mathcal{I}$ of the injection process
in the time interval $[0,T]$ then, by applying \eref{Flow n} as well as the
rules \eref{Rule 1}--\eref{Rule 2} at collisions, one obtains a flow
$\Phi^{t}(\cdot,\mathcal{I})$ on the full state space $\Omega$. Thus, if the
system is in the state $\xi_0 \in \Omega$ at time $t=0$, then its state at any
later time $t \in (0,T]$ is given by
\begin{equation}\label{flow}
\xi(t) \equiv \Phi^{t}(\xi_0,\mathcal{I}) = (\mathbf{q}(t), \phi(t),
\mathbf{v}(t), \omega(t)) \in \Omega\p.
\end{equation}

The scheme described above leaves collisions with the corners
$\partial\Gamma^*$ of the cell $\Gamma$ undetermined. When we discuss
controllability, such
orbits will not be considered.
Similarly,
we shall only consider dynamics so that at most
one particle collides with the disk at any given time. The state space associated to the $N$-cell system will be introduced in \Section~\ref{N-cell analysis}.


\subsection{The strategy}

Here, we outline the strategy adopted to show the controllability of our
class of systems. Note first that the mechanical nature of the class of systems considered in this paper makes them \emph{time-reversible}. Thus, one obtains
controllability of any system in our class by establishing a way to drive
(in a finite time) the system from any state to the ground state,
\ie the state in which there is no particle and all disks have zero
angular positions and zero angular velocities. We shall start with the one-cell
system and easily obtain its controllability from the following three crucial
properties:
\begin{enumerate}
\item[1.] Given an initial state $\xi_0 \in \Omega$, there is a way to set the angular
  velocity and the angle of the disk  to any prescribed value in an
arbitrary
short time (in particular before any particle collides with the
  disk). This operation can be achieved by particles which fly into the
  cell from outside, hit the disk, and exit again (all this before the
  next collision of another particle with the disk).
The particles used for this process exist because of  our assumptions
  on the nature of the heat baths: They will be called 
\emph{drivers}.
\item[2.] Any \emph{admissible path} in the cell (to be  defined) can be
  realized by a particle in
  the system, which we shall call a \emph{tracer}, by controlling its trajectory by acting adequately with driver particles on the disk.

\item[3.] If the cell is 1-controllable, then there exists in fact an
  admissible path 
between any point $\theta_\0$ on the disk and one of the openings
  $\partial\Gamma_{\L}$ or $\partial\Gamma_{\R}$ (one can choose
  which one).
\end{enumerate}
In the $N$-cell situation, we will obtain controllability by generalizing the
strategy described  above.


\section{One-cell analysis}\label{One-cell analysis}

\subsection{Paths of a particle}

In this subsection, we consider one particle
in one cell and
characterize the set of possible paths it can follow (with the help of
other particles) under the
collision rules \eref{Rule 1}--\eref{Rule 2} at $\partial
\Gamma$. We will extend that later in a straightforward way to an
arbitrary number of particles.

\begin{definition}\label{def1}\label{Assumptions One Cell}
A curve $\gamma : s\mapsto
  \gamma(s)\in \Gamma$,
  $s\in [0,1]$, is called an
  \emph{admissible path} if it is continuous on $[0,1]$, piecewise
differentiable on $(0,1)$ and satisfies the
following properties:
\begin{enumerate}
\item[1.] It consists of a finite sequence of straight segments meeting
  at the
boundary $\partial \Gamma = \partial\Gamma_{\outer} \cup \partial D$ of the cell.
\item[2.] The incoming and outgoing angles of two consecutive segments of $\gamma$
meeting on the outer boundary $\partial \Gamma_\outer$ of the cell are equal.
\item[3.] Only its end points $\gamma(0)$ and $\gamma(1)$ can be in the openings $\partial\Gamma_\L$ and $\partial\Gamma_\R$.
\item[4.] It does not meet any corners of the cell, \ie $\gamma(s) \not\in \partial
\Gamma^*$ for all $s \in [0,1]$.
\item[5.] It is nowhere tangent to the boundary of the disk $\partial D$.
\end{enumerate}
\end{definition}

\begin{figure}[htbp]
\hspace{28.5mm}\psfig{file=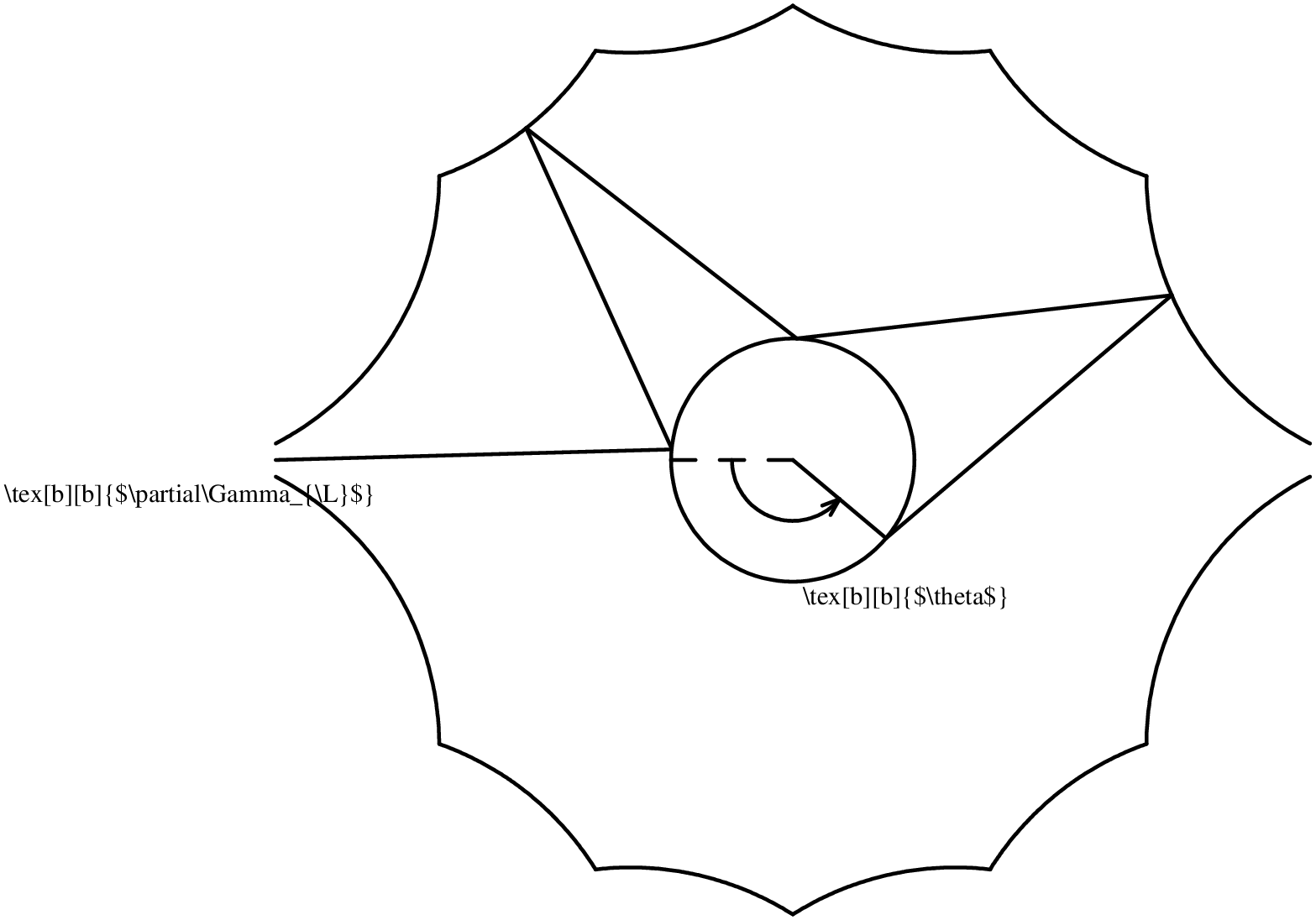,height=4.5cm} \hspace{3cm}
\psfig{file=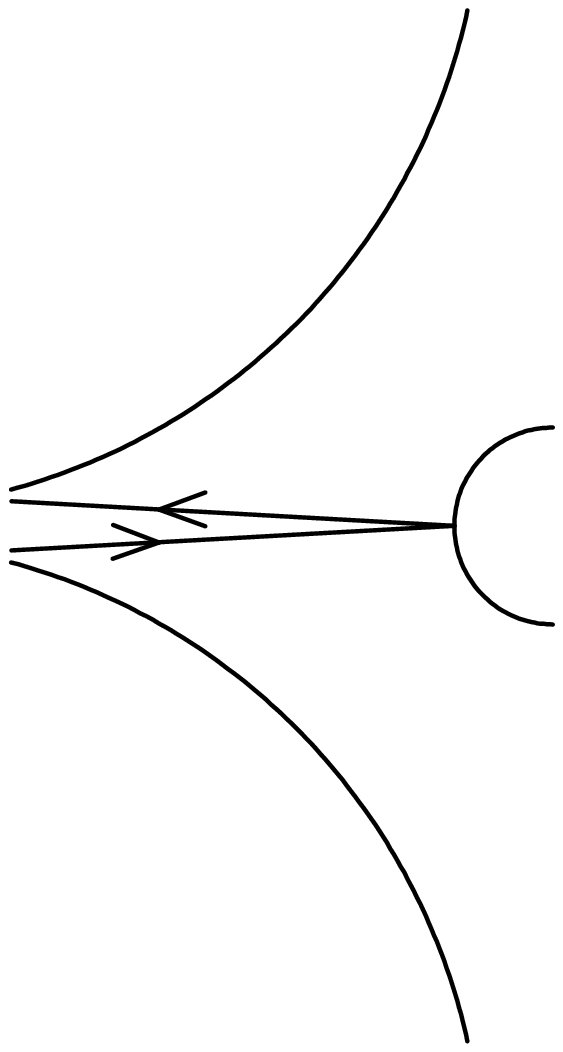,height=4.5cm}
\vspace{3mm}\caption{Left: An admissible path. Right: One possible orbit of the
driver particle.}
\label{Disk Control}
\end{figure}

An example of admissible path is shown in \Fig~\ref{Disk Control}. In the
subsequent development, we shall denote by $|\gamma|$ the length of an
admissible path $\gamma$, \ie $|\gamma| = \sum_{i=0}^{m-1} \int_{s_i}^{s_{i+1}}
|\gamma'(s)| \d s$ if $\gamma$ is made up of $m$ straight segments ($0=s_0 < s_1
< \dots < s_m = 1$).

\begin{remark}
\textnormal{Note that an admissible path
does not need to satisfy any particular ``law of reflection''
on the boundary $\partial D$ of the disk (see \Fig~\ref{Disk Control}).}
\end{remark}

We will show that, by shooting in ``driver'' particles from the
opening $\partial \Gamma_\L$ (or $\partial \Gamma_\R$)
in a well-chosen way, any admissible path can be realized as the
orbit of a ``tracer'' particle moving according to the laws \eref{Rule
1}--\eref{Rule 2} we gave earlier and that
this is possible for any initial speed of the tracer particle (provided it
is strictly positive) and any initial angular velocity of the disk.

We start with the following crucial lemma which
shows that very fast particles coming from the baths can set the disk
to any prescribed angular velocity $\omega$ and leave the system in a
very short time $\delta$. In the sequel, these fast particles will be called \emph{drivers}.

\begin{lemma}\label{Drivers One Cell}\label{driveonedisk} Assume that at time $0$
the disk rotates with angular velocity $\hat\omega$ and that none of the
  particles which are inside the cell will collide with the disk
  before time $\tau > 0$. Then, given any
$\omega \in \real$
and $0 < \delta < \tau$, there exists a way to inject a particle into the cell from
the left entrance $\partial \Gamma_{\L}$ at time $0$
such that at time $\delta$ the disk has angular velocity $\omega$ and the particle has
left
the system (through $\partial \Gamma_\L$). The same holds for
$\partial \Gamma_\R$.
\end{lemma}

\begin{remark}
 \textnormal{The choice of the initial time equal to $0$ is for convenience,
and we
will use the lemma for other initial times as well.}
\end{remark}
\begin{remark}\label{disk at rest}
 \textnormal{Assume we want to describe a strategy which should
 achieve some goal within a lapse of time $\delta$. Then, by \Lemma~\ref{Drivers One Cell}, we can use a
 fraction of this time, say $\delta /2$, to stop the disk, and the other half of the time to do the actual task. So, without loss of
 generality, we may assume that the disk is at rest when the actual
 task begins.}
\end{remark}

\begin{remark}\label{Remark Disk State}
 \textnormal{Note that \Lemma~\ref{Drivers One Cell} actually permits one to
set both the angular velocity $\omega$ \emph{and} the angular position $\phi$
of the disk at time $\delta$. Assume for
illustration that the disk is initially in the state
$(\hat{\phi}=0,\hat{\omega}=0)$ and proceed as follows: send a driver to set
the velocity of the disk to $\omega_1$ at time $\delta_{1} < \delta$ and send a
second driver to set its velocity to $\omega$ at time
$\delta$ such that $\omega_1
(\hat\delta-\hat\delta_{1})/2 + \omega (\delta - \hat\delta/2) = \phi$, where
$\hat\delta_{1}/2$ and $\hat\delta/2$ denote (as in the proof of
\Lemma~\ref{Drivers One Cell}) the collision times of the first and
respectively second driver with the disk.}
\end{remark}

\begin{proof}[{\rm \textbf{Proof of lemma \ref{driveonedisk}}}] To simplify the discussion, we
assume $\hat\omega \geq 0$. Consider the
general setup of \Fig~\ref{The Cell}. The axes are chosen such that the
injection
takes place in the segment $\partial\Gamma_{\L}$ (of length $2a$ and at
$x$-coordinate $0$), the center
of the disk has $y$-coordinate $0$
and has its leftmost point at $(d,0)$. The process we shall realize is sketched
in \Fig~\ref{Disk Control} (the arrows correspond to the case $\omega \geq 0$).
Choose
$\hat{\delta}$ such that
\begin{equation}\label{Condition on delta}
 0 < \hat{\delta} <
\delta \hspace{5mm} \mbox{and} \hspace{5mm} \frac{2
}{\hat{\delta}}
> \max\{\frac{|\omega| }{a}, \frac{\hat{\omega}}{a}\}~.
\end{equation}
Define $v_{x}$ and $\epsilon$ by
\begin{equation}\label{Definition of V and epsilon}
v_{x} = \frac{2 d}{\hat{\delta}} \hspace{5mm} \mbox{and}
\hspace{5mm} \epsilon = \frac{\omega d}{v_{x}}~.
\end{equation}
Clearly, $|\epsilon | <a$.
We inject a particle into the cell at time
$0$ at the point $(0,-\epsilon)$, with velocity
$(v_{x},\omega)$. No other particles are injected in the time interval
$[0, \delta]$.
Before the collision with the disk the particle follows
the path:
\begin{equation*}\{x(t) = v_{x}\, t , \ y(t) = \omega\, t - \epsilon \mbox{ for } t \in
[0,\hat{\tau}]\}\p~,
\end{equation*}
where $\hat{\tau}$ denotes the collision time. By construction,
the particle hits the disk at the point $(d,0)$ at time
$\hat{\tau}=\hat\delta/2$.
At the collision, the tangent velocity of the particle is
exactly $\omega$ and the disk rotates at angular velocity $\hat{\omega}$. After
the collision, the particle has velocity $(-v_{x},\hat{\omega})$ and
follows the path:
\begin{equation*}
\{x(t) = d - v_{x} (t - \hat\tau), \ y(t)= \hat{\omega} (t
- \hat\tau) \mbox{ for } t \in [\hat\tau,2\hat\tau] \}\p~.
\end{equation*}
At time $2 \hat\tau = \hat\delta$, the particle is at
$(0,\tilde{y}=\hat{\omega} \hat\delta/2)$. Since $0 \leq \tilde{y} < a$ by \eref{Condition on
delta} the particle will have
reached $\partial\Gamma_{\L}$ at time $\hat\delta$  and will exit the cell.
Note that if $v_{x}=\hat{\omega} d / a$ then $\tilde{y} = a$, so
that the second condition in \eref{Condition on delta} demands that
the ($x$-component of the) incoming velocity is sufficiently large so
that the
particle will not miss the exit.
\end{proof}

\begin{proposition}\label{Path} Let $\gamma$ be an admissible path
  and assume that a particle starts at time $0$ from $\gamma(0)$ with velocity $v_0\ne 0$ in the positive
  direction along $\gamma$.
Then one can find a sequence of drivers such that the particle will
  follow $\gamma$ to its end in a finite time. In particular, if the end of\/
$\gamma$ is in
  $\partial \Gamma_\L$ or $\partial \Gamma_\R$ the particle will leave the
cell.
\end{proposition}

\begin{proof}[{\rm \textbf{Proof}}] Consider first the case where $\gamma$ does not intersect the
boundary
$\partial D$ of the disk. In this situation the admissible path $\gamma$ is
  automatically followed by the particle, since by \eref{Rule 1} the
reflections on the outer boundary of the cell are
  specular. Moreover, the entire path $\gamma$ is realized in a finite time
$T=|\gamma|/|v_0|$ since the norm of the particle's velocity $|v_0|$ is
conserved at all times and initially non-zero. It thus suffices to discuss the
intersections of the admissible path $\gamma$  with the
  disk. Here, we will use drivers to direct the particle along $\gamma$. It
will become clear that if one can do this for one collision with the disk
  one can do it for any finite number of them.

Assume that $\gamma$ hits $\partial D$ for the first time
at $s_1 \in (0,1)$ and decompose $\gamma$ into two parts: the path
before the intersection $\gamma_{0} := \{
\gamma(s) \ | \ s \in [0,s_1] \}$ and the path after the
intersection $\gamma_{1} := \{\gamma(s) \ | \ s \in [s_1,1]\}$.
Since there are only specular reflections up to time $t_1 = |\gamma_0|/|v_0|$,
the particle will follow the path $\gamma_0$ without driver intervention and
will arrive at the impact point $\gamma(s_1) \in \partial D$ at time $t_1$ with
some velocity $v_{\mathrm{in}}$ satisfying $|v_{\mathrm{in}}| = |v_0|$. Let
$\e_{\n}$ and $\e_{\t}$ be unit vectors, respectively normal (outwards) and
tangent to $\partial D$ at $\gamma(s_1)$, and let us write
$v_{\mathrm{in}}=v_{\n} \e_\n + v_\t \e_{\t}$. Note that $v_\n > 0$. If the disk has angular velocity
$\hat \omega$ at the impact time $t_1$, then, by the collision rule \eref{Rule
2} the particle will leave the disk
with velocity $v_{\mathrm{out}}=-v_{n} \e_{\n} + \hat\omega \e_{\t}$. Let
$\alpha \in
(-\frac{\pi}{2},\frac{\pi}{2})$ be the angle between $v_{\mathrm{out}}$ and
$-\e_{\n}$ (\Fig~\ref{The Cell}). Clearly, one has
\begin{equation}
\alpha = \arctan\left({\hat\omega}/{v_{\n}}\right)~.
\end{equation}
Hence, in order to force the particle
to emerge from the impact point
in any prescribed direction $\alpha$ (which is not tangent to the impact
point), in particular in the direction of $\gamma_1$, it suffices to let a
driver arrive at the disk at time $\tau_1$ before $t_1$ to give the disk the
appropriate angular velocity
$\hat\omega$.

To follow the full path $\gamma$ we proceed by induction over the
intersections with the disk and this concludes the proof. Note that the norm of
the particle's velocity is not conserved along the orbit,
so that the total time $T$ the particle takes to complete the entire path
$\gamma$ is not $|v_0|/|\gamma|$. Note however that because $\gamma$ is nowhere
tangent to the disk the normal component $v_\n$ is non-zero at each collision
so that the total time $T$ is anyhow finite.
\end{proof}

\begin{remark}\label{remcoll}
\textnormal{
The precise details used in \Proposition~\ref{Path} to constrain the
tracer particle along the path $\gamma$ are not
unique. Note first that given an admissible path $\gamma$ and an initial velocity
$v_0$,
the speed of the tracer in each straight segment of $\gamma$ is determined by
the rules \eref{Rule 1}--\eref{Rule 2} of
collision. Therefore, there is a sequence of times
$t_{1} < \dots < t_{m}$ at which the tracer will hit the
disk. The times $\{\tau_{1},\dots,\tau_{m}\}$ at which
the drivers set the angular velocity of the disk to the appropriate value only
have to satisfy
\begin{equation}
\tau_{1} < t_{1} \hspace{5mm} \mbox{and} \hspace{5mm}
t_{i-1} < \tau_{i} < t_{i}\p~.
\end{equation}
Indeed, any sequence $\{\tau_{1},\dots,\tau_{m}\}$ satisfying
these conditions is acceptable in the context of \Proposition~\ref{Path} and
for every $j \in \{1,\dots,m\}$ there exist
infinitely many $\delta_{j} \in (0,t_{j}-\tau_{j})$
that can be considered in \Lemma~\ref{Drivers One Cell}.}
\end{remark}


\subsection{Repatriation of particles}

In this subsection, we use the specific properties of the cell
(\Section~\ref{Geometry of a Cell}) to control the trajectories of the
particles after they have encountered the disk. In particular, the results
established here will be necessary in the $N$-cell analysis to bring back the
drivers from a given cell to one of the baths.

\begin{lemma}\label{Path Exit Disk}
Let $\theta_\0 \in \partial D$ and assume that the cell is $1$-controllable.
Then there exists an admissible path between $\theta_\0$ and
$\partial\Gamma_{\L}$ (or $\partial\Gamma_{\R}$).
\end{lemma}

\begin{remark}\label{remark ergodicity}
\textnormal{Note that ergodicity is not a sufficient condition to obtain the
above result. Indeed, consider the following system: a particle in a cell with closed entrances ($a=0$) and with a circular inner boundary. Assume that all collisions of the particle in the cell are specular. Notice that our
model can be reduced to this system by using the drivers of \Lemma~\ref{Drivers
One Cell} (before each collision with the disk, use a driver to set $\omega = v_t$, where $v = v_{\n} e_\n + v_\t e_\t$ is the velocity of the particle at the collision time; this will mimic a specular reflection). Then, even though it is well known that such a system is ergodic \cite{Sinai,B}, one still
cannot conclude that there exists a trajectory between $\theta_\0$ and
$\partial\Gamma_{\L}$ that does not intersect $\partial\Gamma_{\R}$ in between.
For this one needs to control the trajectory (see the proof below).}
\end{remark}

\begin{proof}[{\rm \textbf{Proof}}]
We shall exploit the properties of the illuminated segments $I_1,\dots,I_b$ (\Section~\ref{Geometry of a Cell}). Consider a point $\theta$ in $I_k$, for some $k \in \{1,\dots,b\}$. A particle leaving this point in
the direction of the center $c_k$ will return to $\theta$ after one collision with
$\partial \Gamma_k$. Clearly, if one changes the direction
sufficiently little, the particle will return to a point $\theta'$ which is still in $I_k$. Consider the union of the open intervals
$(\theta,\theta')$
(respectively $(\theta',\theta)$ if $\theta' < \theta$) obtained in this fashion. Since every illuminated segment is an open connected set, one obtains, by varying the index $k$ over $\{1,\dots,b\}$, an open cover $\mathcal{O}$ of the illuminated region $I=\cup_{k=1}^{b} I_k$.

By assumption of 1-controllability, one has $I = \partial D$ and it follows, by the Heine-Borel theorem, that there exists a
finite subset of $\mathcal{O}$ which covers the entire boundary of the disk.
Therefore one finds, for \emph{any} two points
$\theta_{\rm initial}$ and $\theta_{\rm final}$ on the boundary of the disk, a sequence 
$(\theta_1, \dots, \theta_m)$ of angles, with $\theta_1 = \theta_{\rm
  initial}$ and $\theta_m = \theta_{\rm final}$,
such that an admissible path from $\theta_{\rm initial}$ to $\theta_{\rm
  final}$ can be realized by ``jumping'' from $\theta_i$ to
$\theta_{i+1}$, for $i=1,\dots,m-1$ (each time via some $\partial\Gamma_k$ with a
specular reflection).

Finally, if the orbit has reached an angle from which there
is a direct line joining the left exit (without intersecting the boundary $\partial\Gamma_{\outer}\setminus (\partial\Gamma_{\L}\cup\partial\Gamma_{\R})$),
we choose that line and we are done (see \Fig~\ref{Disk Control}).
\end{proof}

\begin{remark}\label{Many Admissible Paths}
\textnormal{Notice that the set of intermediate points $(\theta_1, \dots, \theta_m)$ between $\theta_{\rm initial}$ and $\theta_{\rm final}$ is open in $\real^m$. It follows that there actually exists an \emph{open} set of admissible paths between a given point $\theta$ on the disk and the left exit
$\partial\Gamma_{\L}$, each of which having different intermediate intersection
points with the disk.}
\end{remark}

\begin{remark}
  \textnormal{While the proof of \Lemma~\ref{Path Exit Disk} uses the Heine-Borel theorem, which in
    its standard form is non-constructive, it is in principle easy for
    any given region to actually invent a constructive proof. For example, one can proceed as follows: Fix any pair of points $\theta_{\rm initial}$ and $\theta_{\rm final}$ in a given illuminated region $I_k$ and determine a \emph{uniform} lower bound $\Delta\theta > 0$ for the displacement of a particle within $[\theta_{\rm initial}, \theta_{\rm final}]$ through specular reflections from $\partial \Gamma_k$. 
Such a uniform bound can be obtained by considering the worst possible situation in $[\theta_{\rm initial}, \theta_{\rm final}]$. This shows that there exists an admissible path between any two points in a given illuminated region. One then concludes, as in the above proof, by using the assumption of 1-controllability. Since the arithmetics is somewhat involved, we omit this construction.
}
\end{remark}
\begin{corollary}\label{Admissible Paths Between Openings}
If the cell is $1$-controllable, then there exists an admissible path between
$\partial\Gamma_{\L}$ and $\partial\Gamma_{\R}$ so that its end points are
located at the center of the straight boundary pieces and its first and last
straight segments are orthogonal to them (see \Fig~\ref{Admissible Paths
Openings}). Furthermore, such a
path exists also for which the first and last straight
segments make a ``small'' angle with the horizontal.
\end{corollary}
\begin{proof}[{\rm \textbf{Proof}}]
  The statements are obvious, by considering the proof of \Lemma~\ref{Path Exit Disk} with the angles $\theta_{\rm
  initial}$ and $\theta_{\rm final}$ corresponding to the
  points where the first and respectively last straight segment intersect the disk.
\end{proof}

\begin{figure}[htbp]
\hspace{32.5mm}\psfig{file=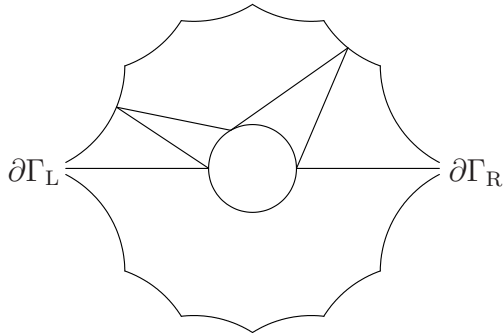,width=5cm}
\vspace{3mm}\caption{An admissible path linking the two openings.}
\label{Admissible Paths Openings}
\end{figure}


\subsection{Orbits of the system}

\noindent We define the \textit{ground state} $\xi_{\g} \in \Omega$
of the system as the state in which the system is empty ($\xi_{\g}
\in \Omega_{0}$) and the disk is at rest ($\omega  = 0$) at zero
angular position ($\theta = 0$). In this subsection, we show that a suitable
realization of the injection process can drive the system from any (admissible)
initial state $\xi_0 \in \Omega$ to the ground state.

\begin{definition}\label{Admissible state}
A state $\xi_0 =
(q_{0,1},\dots,q_{0,n},\phi_0,v_{0,1},\dots,v_{0,n},\omega_0)\in \Omega_n$ is
called an \emph{admissible initial state} (at time $0$)
if it satisfies the following properties ($i,j = 1, \dots, n$):
\begin{enumerate}
\item[1.] The particles are initially inside the cell with non-zero velocities: $q_{0,i} \in \Gamma\setminus\partial\Gamma$ and $v_{0,i} \not = 0$.
\item[2.] The particles will either hit the disk or exit: for each $i$ there is a finite time
$t_{i} > 0$ such that $q_{i}(t_i) \in \partial D \cup \partial\Gamma_{\L} \cup
\partial\Gamma_{\R}$ and  $q_{i}(t) \not\in \partial D \cup \partial\Gamma_{\L}
\cup \partial\Gamma_{\R}$ for $0 < t < t_i$.
\item[3.] No tangent collisions with the disk: if $q_{i}(t_i) \in
\partial D$, then the normal component
$v^{\n}_{i}(t_i)$ of $v_{i}(t_i)$ to $\partial D$ at $q_{i}(t_i)$ is non-zero.
\item[4.] No simultaneous collisions with the disk: if $q_{i}(t_i)
\in \partial D$ and $q_{j}(t_j) \in \partial
D$ with $i \neq j$, then $t_i \neq t_j$. 
\item[5.] No collisions with the corner points of the cell: $q_{i}(t) \not\in
\partial \Gamma^*$ for $0 < t \leq t_i$.
\end{enumerate}
\end{definition}

\begin{remark}\label{Rem on Admissible States}
\textnormal{The second condition in property~1 as well as properties~2 and 3 are necessary to prevent particles from staying forever in the system. (Note that a tangential collision with the disk at rest would stop the particle forever.) The other properties are necessary to get rid of all
undefined events. Using the well-known fact that the cell without the disk
constitutes an ergodic system \cite{Sinai,B}, one easily sees that the set of states in
$\Omega_n$ which do not satisfy these properties is negligible with respect to Liouville measure.}
\end{remark}

\begin{definition}\label{Assumptions One Cell - n particles}
An \emph{admissible movie} is a set of $n$ admissible paths $\gamma_1,\dots,
\gamma_n$ each of which being equipped with a tracer initially located at
$\gamma_i(0)$ with velocity $\overline{v}_i(0)$ directed positively along
$\gamma_i$ such that
\begin{enumerate}
\parskip -3pt
\item[1.] Each  $\gamma_i$ ends at the exits: $\gamma_i(1) \in \partial\Gamma_{\L}
\cup \partial\Gamma_{\R}$.
\item[2.] Each tracer follows its corresponding admissible path up to the end in a
finite time.
\item[3.] The scattering events on the disk are not simultaneous.
\end{enumerate}
\end{definition}

\begin{theorem}\label{Admissible Movie Existence}
Let $\xi_0 \in \Omega_n$ be an admissible initial state and assume the cell to
be $1$-controllable. Then there exists an admissible movie with $\gamma_i(0) =
q_{0,i}$ and $\overline{v}_i(0) = v_{0,i}$\, for $i=1,\dots,n$.
\end{theorem}

\begin{proof}[{\rm \textbf{Proof}}]
Let us put a tracer at each position $q_{0,i}$ with velocity $\overline{v}_i(0)
= v_{0,i}$ for $i=1,\dots, n$. Then, by \Definition~\ref{Admissible state},
there exist finite times $t_{i} > 0$ ($1 \leq i \leq n$) at which each tracer
either leaves
the cell (without making any collision with the disk) or hits the disk:

(a) If the $i$th tracer is in the first alternative, we consider its path
$\gamma_i = \{q_i(t)\ | \ t\in [0,t_i]\}$ which is clearly admissible.

(b) In the second alternative, we denote by $\gamma_{i}^{-}$ the path realized
by the
$i$th tracer between time $0$ and the collision time $t_{i}$ (along which there
is no collision with the disk). By \Lemma~\ref{Path Exit Disk}, there exists
an admissible path $\gamma_{i}^{+}$
between the collision point on the disk and the left exit. We then consider the
following admissible path: $\gamma_i = \gamma_{i}^{-} \cup \gamma_{i}^{+}$.

We denote by $\mathcal{C} \subset \{1,\dots,b\}$ the set of subscripts corresponding to the particles which are in case (b). 
Then, by \Proposition~\ref{Path} combined with \Remarks~\ref{remcoll} and \ref{Many Admissible Paths}, one can
choose the admissible paths $\gamma_{j}^{+}$ ($j \in \mathcal{C}$) and inject the
drivers that are used to constrain the $j$th tracer along $\gamma_{j}^{+}$ in
such a way that all drivers and tracers involved in the movie do not make any
simultaneous collisions with the disk. More precisely, there exist
admissible paths and a set of drivers so that
the tracers will hit the disk at distinct times $\tau_1 < \dots < \tau_m$ and the drivers will be in the system 
only in the time intervals $[\tau_i,\tau_{i+1})$, for $i = 1,\dots,m-1$, during each of which they control the disk 
in such a way that the tracer leaving the disk at time $\tau_{i+1}$ has the appropriate direction. This ends the proof.
\end{proof}

Taking into account \Remarks~\ref{Remark Disk State} and \ref{Rem on Admissible
States} as well as \Remarks~\ref{remcoll} and \ref{Many Admissible Paths} one obtains the following result as a consequence of the preceding
theorem:

\begin{corollary}
Assume the cell to be 1-controllable. Then, for almost every initial state
$\xi_0 \in  \Omega$ (with respect to Liouville) there exist a finite time $T > 0$ and an open set $\mathcal{B}([0,T])$ of realizations
of the injection process in the time interval $[0,T]$ such that $\Phi^T(\xi_0,\mathcal{I}) = \xi_\g$  for all $\mathcal{I} \in \mathcal{B}([0,T])$.
\end{corollary}


\section{N-cell analysis}\label{N-cell analysis}

We now extend the preceding results to the $N$-cell system. For this we need to
introduce the corresponding notations and terminologies.

A system composed of $N$ identical $1$-controllable cells is said to be
$1$-controllable. A continuous path in the system which is composed of finitely
many admissible paths is also called an admissible path. The particles that
will be used to control the angular velocity of a given disk in the system will
still be called drivers and those which will follow admissible paths will again
be called tracers.

We write $\Gamma^{N}=\Gamma^{(1)} \times \dots \times \Gamma^{(N)}$ for the
domain accessible to the particles in the system composed of $N$ identical
cells, where each $\Gamma^{(\ell)}=\Gamma_{\outer} ^{(\ell)}\setminus D^{(\ell)}$ can be identified with
$\Gamma$, and denote by $\Omega^{N}= \bigcup_{\ell=1}^{N}
\bigcup_{n=0}^{\infty}\Omega^{(\ell)}_{n}$ the corresponding state space, where
each $\Omega^{(\ell)}_{n}$ is defined as
in \eref{State Space with n particles}. We also define
$\Omega^{N}_{n_1,\dots,n_N}=\Omega^{(1)}_{n_1} \times \dots  \times
\Omega^{(N)}_{n_N}$ so that
$\Omega^{N}=\cup_{n_1,\dots,n_N = 0}^{\infty} \Omega^{N}_{n_1,\dots,n_N}$. A state  $\xi \in \Omega^{N}_{n_1,\dots,n_N}$ is written as follows:
\begin{equation}\label{State Form}
\xi = (q_{1},\dots,q_{n},
\phi_{1},\dots,\phi_N, v_{1},\dots,v_{n},\omega_{1},\dots,\omega_N)~,
\end{equation}
where the total number of particles within the system is $n=n_1 + \dots + n_N$. As in \eref{flow} we denote by $\Phi^t(\cdot,\mathcal{I})$ the flow on $\Omega^{N}$. Note that the openings corresponding to the baths are now
$\partial\Gamma^{(1)}_{\L}$ and $\partial\Gamma^{(N)}_{\R}$. Clearly, the
notions of ground state  $\xi_{\g} \in \Omega^N$ and that of admissible movie
can be generalized in a straightforward way to the $N$-cell system. Finally, 
the notion of admissible initial state, given in \Definition~\ref{Admissible
state}, is generalized as follows:

\begin{definition}\label{Admissible state II}
A state $\xi_0 \in \Omega^{N}_{n_1,\dots,n_N}$, written as in \eref{State
Form}, is called an \emph{admissible initial state} if it satisfies the following properties ($\ell, \ell' = 1,\dots,N$ and $i,j = 1,
\dots, n$):
\begin{enumerate}
\item[1.] The particles are initially inside the system with non-zero velocities: $q_{0,i} \in \Gamma\setminus\partial\Gamma$ and $v_{0,i} \not = 0$.
\item[2.] The particles will either hit a disk or exit the system: for each $i$ there is a finite time $t_{i} > 0$ and an index $\ell$ such that $q_{i}(t_i)
\in \partial D^{(\ell)} \cup \partial\Gamma^{(1)}_{\L} \cup
\partial\Gamma^{(N)}_{\R}$ and  $q_{i}(t) \not\in \partial D^{(1)} \cup \dots \cup \partial D^{(N)}
\cup \partial\Gamma^{(1)}_{\L} \cup \partial\Gamma^{(N)}_{\R}$ for $0 < t
< t_i$.
\item[3.] No tangent collisions with the disks: if
$q_{i}(t_i) \in
\partial D^{(\ell)}$, then the normal
component $v^{\n}_{i}(t_i)$ of $v_{i}(t_i)$ to $\partial
D^{(\ell)}$ at $q_{i}(t_i)$ is non-zero.
\item[4.] No simultaneous collisions with the disks: if $q_{i}(t_i)
\in \partial D^{(\ell)}$ and $q_{j}(t_j) \in \partial
D^{(\ell')}$ with $i\neq j$, then $t_i \neq t_j$.
\item[5.] No collisions with the corner points of the system:  $q_{i}(t)
\not\in
\partial \Gamma^{N,*}$ for $0 < t \leq t_i$.
\end{enumerate}
\end{definition}

\begin{remark}
 \textnormal{Note that property 4 excludes simultaneous collisions with any given disk ($\ell = \ell'$), which is necessary since such events are undefined, but it also excludes simultaneous collisions of particles with different disks ($\ell \not= \ell'$). This requirement is actually not necessary but, since such events are negligible (with respect to Liouville), we decided for a matter of convenience to exclude them.}
\end{remark} 

From \Lemma~\ref{Path Exit Disk} and \Corollary~\ref{Admissible Paths Between
Openings} one immediately obtains the following generalized result:

\begin{lemma}\label{Admissible Paths Between Openings II}
Let $\theta_j \in \partial D^{(j)}$ for some $1 \leq j \leq N$ and assume that the
system is $1$-controllable. Then there exists an admissible path between
$\theta_j$ and $\partial \Gamma_{\L}^{(1)}$ (or $\partial
\Gamma_{\R}^{(N)}$).
\end{lemma}

Let us now generalize the second crucial result, namely \Lemma~\ref{Drivers One
Cell}. We want to achieve the controlling of disk $j$ in a very short
time. Basically, one should think that one wants to control disk $j$
\emph{before} some time when a particle hits it, but this controlling should
happen \emph{after} any collision of any other particle with one of the disks
$1,\dots,j-1$. 

\begin{proposition}\label{Drivers One Cell II} 
Assume that the system is $1$-controllable and that at time $0$
the disks rotate with angular
velocities $\hat{\omega}_{1},\dots,\hat{\omega}_{N}$ and that none of the
  particles which are inside the system will collide with any disk
  before time~$\tau > 0$. Then, given $j \in
\{1,\dots,N\}$, $\omega_j \in \real$ and $0 < \delta < \tau$, there exists a way
to inject drivers from the left entrance $\partial\Gamma_{\L}^{(1)}$ at time
$0$
such that at time $\delta$ the $\ell$th disk has angular velocity $\hat{\omega}_{\ell}$
if $\ell \not = j$ and $\omega_j$ if $\ell = j$ and all the drivers have left
the system (through $\partial\Gamma_{\L}^{(1)}$). The same holds for
$\partial\Gamma_{\R}^{(N)}$.
\end{proposition}

\begin{proof}[{\rm \textbf{Proof}}]
The proof is by induction over the subscript $j=1,\dots,N$. The case $j=1$ has
already been treated in the preceding section (\Lemma~\ref{Drivers One Cell}).
Assume now that $j > 1$ and that one can control disks~$1$ to $j-1$. We shall show that there
exists a way to control disk~$j$. Since, by the inductive
hypothesis, one can set the angular velocities of the disks $1,\dots,j-1$ to any values in an arbitrarily short time, one can assume, without loss of generality, that these disks are initially at rest, \ie $\hat{\omega}_{1}= \dots =\hat{\omega}_{j-1} = 0$ (see also \Remark~\ref{disk at rest}). 

As in the proof of \Lemma~\ref{Drivers One Cell} we shall construct a class of
admissible paths $\gamma_j$, with parameters
$(\hat{\omega}_j,\omega_j,\delta)$, starting from the left bath
$\partial\Gamma_{\L}^{(1)}$, going to disk~$j$ and then returning to the left
bath. We shall denote by $\gamma_{\IN}$ the incoming path linking the left bath
to disk~$j$ and by $\gamma_{\OUT}$ the outgoing path from disk~$j$ to the left
bath; thus $\gamma_j = \gamma_{\IN} \cup \gamma_{\OUT}$ (see
\Fig~\ref{Admissible Paths 3 cells}). 

\begin{figure}[htbp]
\hspace{32mm}\psfig{file=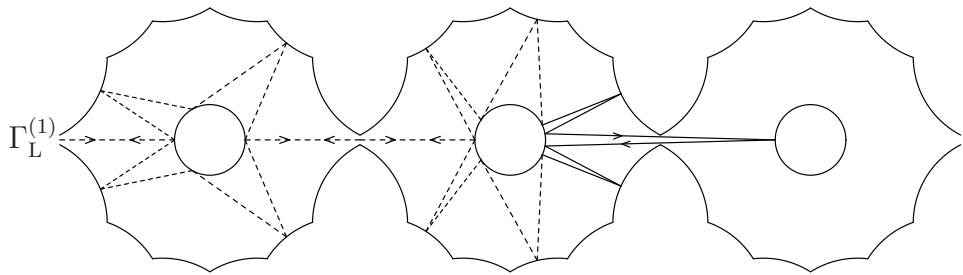,width=12cm}
\vspace{3mm}\caption{The admissible incoming and outgoing paths in the case
$j=3$ ($\hat{\omega}_j \leq 0$, $\omega_j \leq 0$): $\gamma_{\IN}$ is the upper
path and $\gamma_{\OUT}$ the lower one.} 
\label{Admissible Paths 3 cells}
\end{figure}

Consider \Fig~\ref{Parameters gamma out}. We first choose an open segment $\Delta$ centered at $\theta_0$ such that for every $\theta_\IN \in \Delta$ the line emerging from $\theta_\IN$ and intersecting disk~$j$ at the horizontal broken-line does not cross a wall (\ie the boundary $\partial\Gamma_{\outer}\setminus (\partial \Gamma_{\L} \cup \partial\Gamma_{\R})$). For every angle $\theta_\IN \in \Delta$ we choose an admissible path from $\partial\Gamma_\L^{(1)}$ 
to $\theta_\IN$, which exists by \Lemma~\ref{Admissible Paths Between Openings II}. This specifies the incoming path $\gamma_\IN$ (see \Figs~\ref{Admissible Paths 3 cells} and \ref{Parameters gamma out}). We next drive a particle (called the
\emph{controller}) along the incoming path, \emph{where it will play the role of a driver for disk
  $j$}. Given the inductive hypothesis and \Proposition~\ref{Path}, there
is clearly a set of drivers which will drive the controller along this path. We now scale the initial velocities of the controller and of all the drivers by a common factor $\lambda$ and scale the injection times by $1/\lambda$. Note that this scaling preserves the trajectories of the controller and of the drivers.

Similarly, given $\gamma_{\IN}$, $\lambda$ and $\hat{\omega}_j$, there are an associated admissible outgoing path $\gamma_{\OUT}$ (specified by an angle $\theta_\OUT \in \Delta$) and a corresponding sequence of drivers so that the controller will be driven back to the left bath after it has collided with disk~$j$ (provided $\lambda$ is large enough, see below). A typical scenario is shown in \Fig~\ref{Admissible Paths 3 cells}.

\begin{figure}[htbp]
\hspace{25mm}\psfig{file=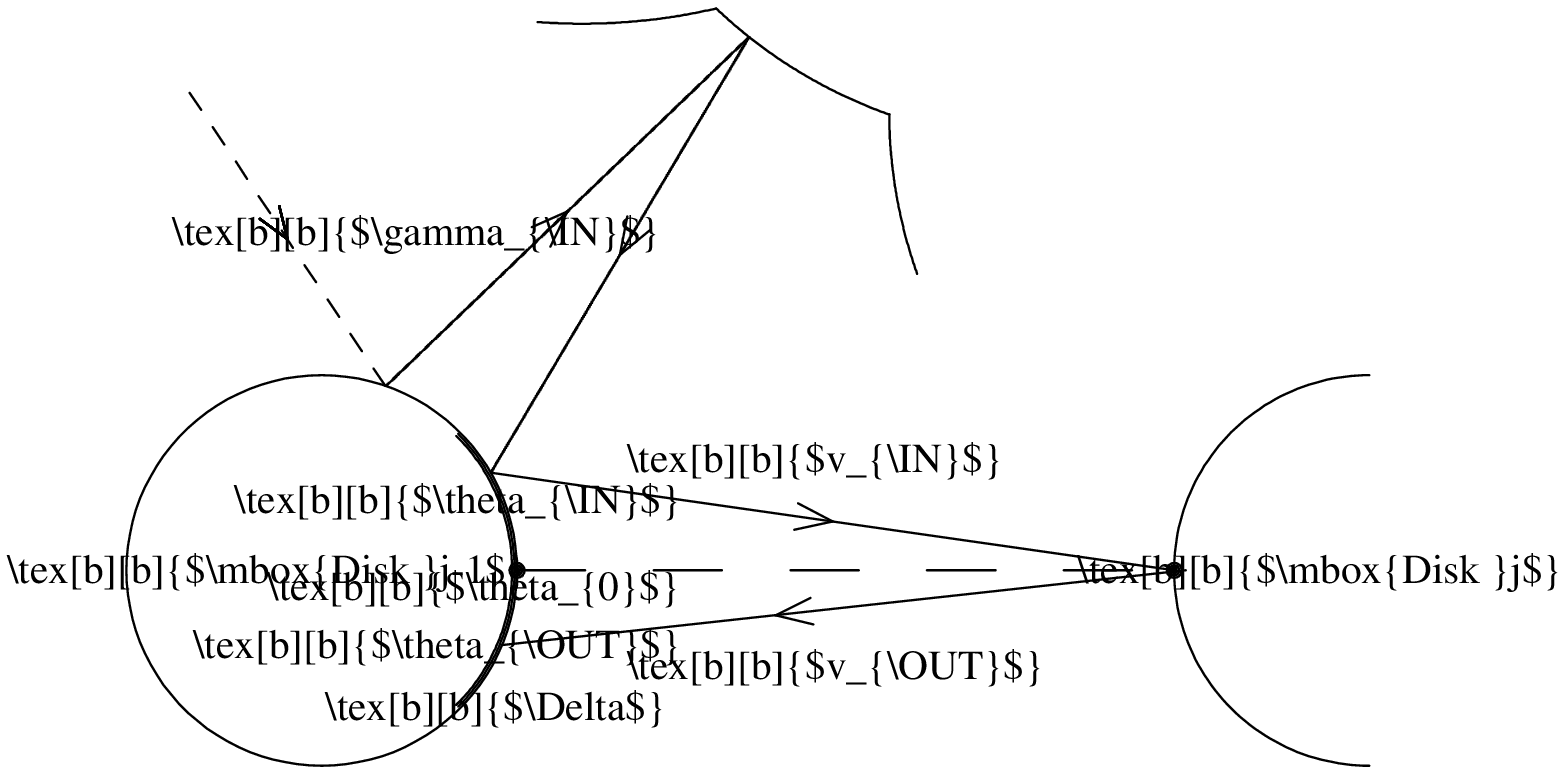,width=9.5cm}
\vspace{3mm}\caption{Some parameters.}
\label{Parameters gamma out}
\end{figure}

 It is clear that one can choose the families of paths $\{\gamma_\IN\}_{\theta_\IN \in \Delta}$ and $\{\gamma_\OUT\}_{\theta_\OUT \in \Delta}$ such that the following properties hold:
\begin{itemize}
\item[1.] The length of the full paths $\gamma_j = \gamma_\IN \cup \gamma_\OUT$ is bounded uniformly in $\theta_\IN, \theta_\OUT \in \Delta$.
\item[2.] For each $\lambda$, the incoming speed $|v_\IN|$ varies continuously with $\theta_\IN$.
\item[3.] For each $\theta_\IN \in \Delta$, the speed $|v_\IN|$ is an increasing and continuous function of $\lambda$. 
\end{itemize}

\noindent{\bf Step 1:} Let  $0 < \delta < \tau$ and $\hat{\omega}_j \in \real$ be fixed. By property~$1$ there is a finite threshold $\lambda_1$ so that, for every $\lambda > \lambda_1$ and every $\theta_\IN \in \Delta$, the controller will travel through $\gamma_{\IN}$, collide with disk~$j$ and return to the left bath through $\gamma_{\OUT}$ in a time shorter than $\delta/2$. Note that, if the initial angular speed $|\hat{\omega}_j|$ of disk~$j$ is big, then $\lambda$ has to be large enough so that the controller will not meet a wall when returning to disk~$j-1$ after its collision with disk~$j$.

To obtain the above statement, one can proceed as follows. First define
$$
T_\IN(\lambda) = \sup_{\theta_\IN \in \Delta} \{\mbox{Time the controller takes to complete } \gamma_\IN \mbox{ starting with speed }\lambda\}\p,
$$
$$
T_\OUT(\lambda) = \sup_{\theta_\OUT \in \Delta} \{\mbox{Time the controller takes to complete } \gamma_\OUT \mbox{ starting with speed } v^*(\lambda)\}\p,
$$
where $v^*(\lambda) = \inf_{\theta_\IN \in \Delta} \{|v_\OUT(\theta_\IN, \lambda, \hat{\omega}_j)|\}$ ($\hat{\omega}_j$ is fixed) if there is a return $\theta_\OUT \in \Delta$ associated to each $\theta_\IN \in \Delta$, and  $v^*(\lambda) = 0$ otherwise. Then, by property~$1$, there is a threshold $0 < \lambda_1 < \infty$ such that the times $T_\IN(\lambda)$ and $T_\OUT(\lambda)$ are finite for all $\lambda > \lambda_1$. Moreover, these traveling times decrease with $\lambda$. Notice finally that for each $\theta_\IN \in \Delta$ the traveling time of the controller along the full path $\gamma_j = \gamma_\IN \cup \gamma_\OUT$ is bounded by $T_\IN(\lambda) + T_\OUT(\lambda)$.

\noindent{\bf Step 2:} Let $\omega_j \in \real$ be given. From the properties $2$ and $3$ it follows that one can choose $\lambda > \lambda_1$ and the angle $\theta_\IN \in \Delta$ so that disk~$j$ will have the required angular velocity after the controller has collided with it. Note that if one wants
to give a very small angular velocity to disk $j$, it suffices to
choose $\theta_\IN$ sufficiently close to $\theta_0$. 

\noindent{\bf Step 3:} In the remaining time $\delta/2$ we stop the disks~$1$ to $j-1$.

Therefore, by choosing $\lambda$ sufficiently large and the angle $\theta_\IN$ correctly, the disk~$j$ will have any required angular velocity at time $\delta$, the controller (and all drivers) will have left the system and all the perturbed disks (with subscript smaller than $j$) will have been restored to their initial state. 
\end{proof}

Finally, using
\Proposition~\ref{Drivers One Cell II}, one obtains by inspection of the proof of \Theorem~\ref{Admissible Movie
Existence} the main result:

\begin{theorem}\label{main}
Assume the system to be $1$-controllable. Then, for every admissible initial
state $\xi_0 \in \Omega^{N}_{n_1,\dots,n_N}$ there exists an admissible movie
with $\gamma_i(0) = q_{0,i}$ and $\overline{v}_i(0) =
v_{0,i}$ for $i=1,\dots,n$. In particular, for almost every initial state $\xi_0 \in \Omega^{N}$ (with respect
to Liouville) there exist a finite time $T > 0$ and an open set $\mathcal{B}([0,T])$ of realizations
of the injection process in the time interval $[0,T]$ such that $\Phi^T(\xi_0,\mathcal{I}) = \xi_\g$  for all $\mathcal{I} \in \mathcal{B}([0,T])$.
\end{theorem}

\begin{remark}
p \textnormal{In \Theorem~\ref{main}, we used the notion of admissible movie to show that the system can be emptied of any particle in a finite time. There is another way to obtain this result. Assume that one can control all disks as stated in \Proposition~\ref{Drivers One Cell II}. Then, one can control them so that the particles make \emph{specular reflections} with the disks (see also \Remark~\ref{remark ergodicity}). Since such a system is ergodic \cite{Sinai,B}, there must be a finite time at which the system will be empty. Note that if one can show that the $N$-cell system, \emph{with rotating disks}, is ergodic then one obtains controllability as an immediate consequence.}
\end{remark}

\begin{remark}\label{unicity measure}
 \textnormal{First note that the particles and the disks evolve under deterministic rules and thus the considered systems constitute Markov processes. If one can prove that for a 1-controllable system (composed of $N$ cells) there exists an invariant measure on $\Omega^N$ and that this invariant measure is sufficiently regular, then it follows from controllability (\Theorem~\ref{main}) that it is unique and therefore ergodic. (Time-reversibility and \Theorem~\ref{main} imply that for almost every state $\xi \in \Omega^N$ (with respect to Liouville) and any open set $A \subset \Omega^N$ there is a finite time $T > 0$ such that the probability for the system initially in the state $\xi$ to be inside $A$ after time $T$ is positive:  $P_T(\xi,A) > 0$.)}
\end{remark}


\section{Concluding remarks}\label{Conclusion}

\begin{figure}[htbp]
\hspace{25mm}\psfig{file=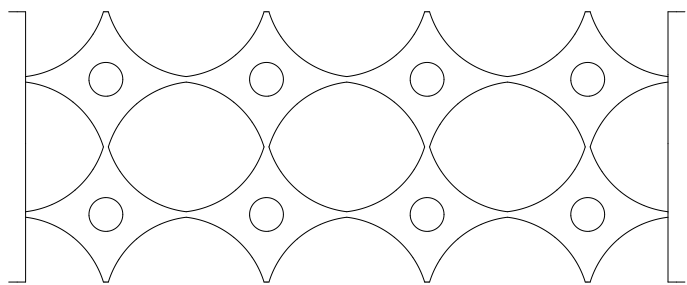,width=9cm}
\vspace{3mm}\caption{A 1-controllable system in 2-d.}
\label{2d controllable}
\end{figure}

We have shown that every chain of 1-controllable identical chaotic cells is controllable with respect to generic baths. As a consequence, one obtains the existence of at most one \emph{regular} invariant measure. The 1-controllable property, introduced through the notion of illumination, allows for a large class of cells and is a rather simple geometrical criterion to check. For the sake of convenience, we have made some simplifying assumptions on the outer boundary $\partial\Gamma_\outer$ of the cell (\ie conditions 1 to 3 in \Section~\ref{Geometry of a Cell}). These assumptions are clearly not optimal to obtain controllability. For example, one can handle systems in which there are some intersection points between $\partial\Gamma_\outer$ and $\partial D$
and in which there are more than one intersection point between the segment $[c,z]$ and $\partial\Gamma_\outer$. However, such a gain of generality was not of interest to us. One could also consider chains of non-identical 1-controllable cells, change the position of the disk or replace it by a some kind of ``potato'' or a needle. One should then be able to control these dynamical scatterers and thus obtain controllability. Note that the present results prove also the controllability for some class of 2-d models; see for example \Fig~\ref{2d controllable}.

\vspace{-6mm}

\ack

The authors thank M.~Hairer,  J.~Jacquet, C.~Mej\'{i}a-Monasterio,
L.~Rey-Bellet, and E.~Zabey for helpful discussions. This work was
partially supported by the Fonds National Suisse.

\section*{References}

\bibliographystyle{unsrt}
\bibliography{refs}

\end{document}